\documentclass{aa}
\usepackage{graphicx}
\usepackage{siunitx}
\usepackage{txfonts}
\usepackage{natbib,twoopt}
\usepackage{comment}
\usepackage[breaklinks,unicode=true,colorlinks,linkcolor=red,citecolor=red,pdfdisplaydoctitle]{hyperref} %% to avoid \citeads line fills
\bibpunct{(}{)}{;}{a}{}{,}             %% natbib format for A&A and ApJ
\newcommand{\parderi}[2]{\frac{\partial #1}{\partial #2}}
\newcommand{\parderii}[2]{\frac{\partial^2 #1}{\partial #2^2}}
\makeatletter
  \newcommandtwoopt{\citeads}[3][][]{\href{http://adsabs.harvard.edu/abs/#3}%
    {\def\hyper@linkstart##1##2{}%
     \let\hyper@linkend\@empty\citealp[#1][#2]{#3}}}
  \newcommandtwoopt{\citepads}[3][][]{\href{http://adsabs.harvard.edu/abs/#3}%
    {\def\hyper@linkstart##1##2{}%
     \let\hyper@linkend\@empty\citep[#1][#2]{#3}}}
  \newcommandtwoopt{\citetads}[3][][]{\href{http://adsabs.harvard.edu/abs/#3}%
    {\def\hyper@linkstart##1##2{}%
     \let\hyper@linkend\@empty\citet[#1][#2]{#3}}}
  \newcommandtwoopt{\citeyearads}[3][][]%
    {\href{http://adsabs.harvard.edu/abs/#3}
    {\def\hyper@linkstart##1##2{}%
     \let\hyper@linkend\@empty\citeyear[#1][#2]{#3}}}
\makeatother

\begin{document} 

  \title{Geminids are initially cracked by atmospheric thermal stress}
  \titlerunning{Geminids are initially cracked by atmospheric thermal stress}

  \author{Tom\'a\v{s} Henych\fnmsep\thanks{Corresponding author, email: ftom@physics.muni.cz}
     \and
     Ji\v{r}\'i Borovi\v{c}ka
     \and
     David \v{C}apek
     \and
     Vlastimil Voj\'a\v{c}ek
     \and
     Pavel Spurn\'y
     \and
     Pavel Koten
     \and
     Luk\'a\v{s} Shrben\'y
   }
  \institute{Astronomical Institute, Czech Academy of Sciences, Fri\v{c}ova 298, 251 65 Ond\v{r}ejov, Czech Republic}
  
   \date{Received 15 October 2025; accepted 4 December 2025}
 
   \abstract
   {Geminids have the highest bulk density of all major meteor showers and their mechanical strength appears to depend on their mass. They are also the most active annual shower, enabling detailed studies of the dependence of their physical and mechanical properties on mass.}
   {We calculated the fragmentation cascades of 39 bright Geminid fireballs, as well as faint video meteors, to derive fragmentation pressures and other physical properties characterizing the meteoroids, such as their bulk densities. Our goal is to describe the mechanical properties across a broad range of initial masses and explain the cause of the observed behavior.}
   {We used a physical fragmentation model with a semiautomatic method based on parallel genetic algorithms to fit the radiometric and regular light curve and dynamics data. We also calculated the thermal stress of model bodies with the type of physical properties and trajectories as the observed Geminids. Then, we compared the outcomes of these simulations to our observations.}
   {We find that the Geminids are probably cracked by thermal stress in the atmosphere first and then eroded by mechanical forces. The most compact Geminids are in the $20\rm{-}200\,\rm{g}$ mass range. The largest observed meteoroids have a wide range of grain sizes, from about $20\,\rm{\mu m}$ to large, non-fragmenting parts of $1\rm{-}20\,\rm{mm}$ in size. The derived bulk densities range from about 1400 to $2800\,\rm{kg\,m^{-3}}$ for smaller meteoroids and approach the assumed grain density of $3000\,\rm{kg\,m^{-3}}$ for larger Geminids.}
   {}

   \maketitle

\section{Introduction}\label{intro}
Geminids are recognized as an amply observed meteor shower, whose physical properties are distinct from the majority of other shower meteoroids \citep{neslusan2015}. Moreover, their parent body, the active asteroid (3200)~Phaethon, is the target of the $\rm{DESTINY}^+$ mission of the Japan Aerospace Exploration Agency (JAXA), aimed at detecting interplanetary dust related to the Geminids.

In \citet{henych2024}, we presented an in-depth study of the mechanical properties of Geminid meteoroids, including statistics on the minimum, mean, and maximum fragmentation pressures for nine modeled Geminid meteoroids (see Fig.~13 of that study). The dynamic pressure of the initial fragmentation seems to depend on the initial meteoroid mass; however, the sample of Geminids in \citet{henych2024} is very limited. Therefore, we set out to expand the sample to include more fireballs over a broader range of initial masses and fainter meteors observed by intensified video cameras. Our main motivation is to verify the robustness of the observed trend and determine the range of masses over which it can be observed. Additionally, more data might allow us to determine the possible cause for this trend.

Second, we aim to describe the mechanical properties of meteoroids accross the widest possible mass range and potentially uncover hidden trends or clustering of these properties. Our primary data source is the systematic observation program of the European Fireball Network~\citep[EN;][]{spurny2007,spurny2017}. However, throughout the year, dedicated observation campaigns of major meteor showers and of regular increased activity of minor meteor showers are carried out, along with studies dedicated to predicted outbursts. These observations typically focus on fainter video meteors and faint fireballs. Therefore, they conveniently supplement the sample of fireballs observed by the EN.

To investigate possible causes of the observed trends, we also modeled the thermal stress induced by the passage of meteoroids through the Earth atmosphere. Thermal stress in meteoroids has been investigated in previous studies. \citet{jones1966} showed that stresses due to the thermal shock in solid stony meteoroids with a radius greater than about $1\,\rm{mm}$ are sufficient to cause fragmentation before ablation begins. They assumed a sudden rise in surface temperature to draw their conclusions. \citet{elford1999} used a more elaborate calculation than \citet{jones1966} and concluded that the fragmentation occurs before ablation even for meteoroids with a radius of about $0.2\,\rm{mm}$ at a speed of $40\,\rm{km\,s^{-1}}$. \citet{tirskiy2004} considered both mechanical and thermal stresses generated in meteoroids and were able to explain the existence of terminal flares observed in bolides.

Thermal stresses generated in meteoroids while still in outer space were modeled by \citet{capek2010}. These authors predicted the faster destruction of larger and weaker Geminid meteoroids. \cite{capek2012} developed a more advanced model whereby thermal stresses create an insulating layer of material that prevents further fragmentation of meteoroids.

In this study, we present the results of fragmentation modeling of Geminid fireballs and faint meteors with a substantially larger sample size than in \citet{henych2024}. Moreover, we calculated the thermal stress in meteoroids generated in the Earth atmosphere. The model bodies had the masses, velocities, and trajectory slopes of the observed Geminids, coupled with the material properties of a carbonaceous material that is plausibly akin to that of Geminids.

In Sect.~\ref{data_mod}, we describe the data and the physical model we used for fitting them. In Sect.~\ref{thermal}, we detail the thermal stress model. We present the results of the study in Sect.~\ref{results} and provide our interpretations  in Sect.~\ref{discussion}. Our conclusions are given in Sect.~\ref{conclusions}.

\section{Data and the modeling procedure}\label{data_mod}
\subsection{Data}
The dataset in the current study comprises 39 Geminids, including 9 taken from \citet{henych2024}. These objects were observed between 2018 and 2023 using various techniques. A list of these meteors, along with their basic physical properties and designations used in some plots, is given in Table~\ref{gem_tab1} (see Sect.~\ref{results}). The initial mass of the meteoroids ranges from $5.7\times10^{-6}\,\rm{kg}$ to $1.7\,\rm{kg}$.

\begin{figure*}
  \centering
   \includegraphics[width=\hsize]{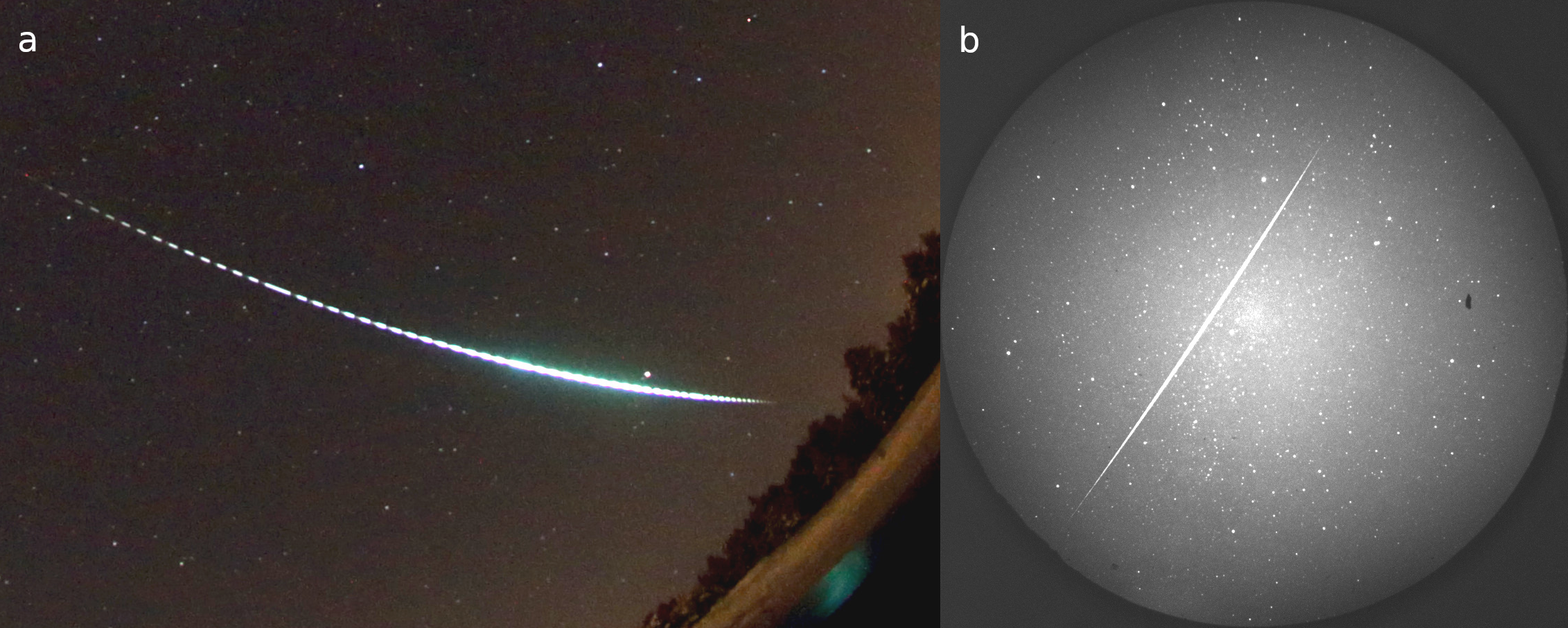}
    \caption{Illustration image of Geminids we used for fragmentation modeling. Panel \textbf{a}: Crop of the all-sky image of the most massive modeled fireball (Geminid~5, \v{S}indelov\'a station). Panel \textbf{b}: Stacked image of the narrow-field-of-view video observation of a faint Geminid 21 (Kun\v{z}ak station).}
    \label{gem_obs}
\end{figure*}

All of the modeled fireballs were observed by the EN, the data reduction process was described in \citet{borovicka2022}. For modeling, the critical factor is the high-time-resolution radiometric light curve of the brightest part of the fireball. The initial and final parts of the composite light curve are captured well by all-sky image photometry and, in some cases, also by security Internet Protocol cameras ($2688\times1520\,\rm{pixels}$ and 20 or 25 frames per second (fps), \citet{borovicka2019,shrbeny2020}). The dynamics of the foremost fragment are also derived from the all-sky images and the Internet Protocol cameras.

Fainter fireballs and faint meteors were observed using two systems: a digital DMK 23G445 GigE monochrome camera coupled with a Mullard XX1332 image intensifier and a Canon 2.0/135~mm lens ($1280\times960\,\rm{pixels}$, 30~fps, field of view of $\sim\!\!22^\circ$, \citealt{koten2020}); and a MAIA camera (JAI~CM-040 CCD camera with a Pentax 1.8/50~mm lens and a Mullard XX1332 image intensifier, $776\times582\,\rm{pixels}$, 61.15~fps, field of view of $\sim\!\!52^\circ$, \citealt{koten2011}) during a dedicated observing campaign in 2022. The observation technique and data reduction for faint meteors were described in detail in \citet{koten2020} and \citet{koten2023}. Figure~\ref{gem_obs} shows an illustrative image of a Geminid fireball taken by an all-sky camera and a faint Geminid meteor captured by an intensified video camera.

\begin{figure}
  \centering
   \includegraphics[width=\hsize]{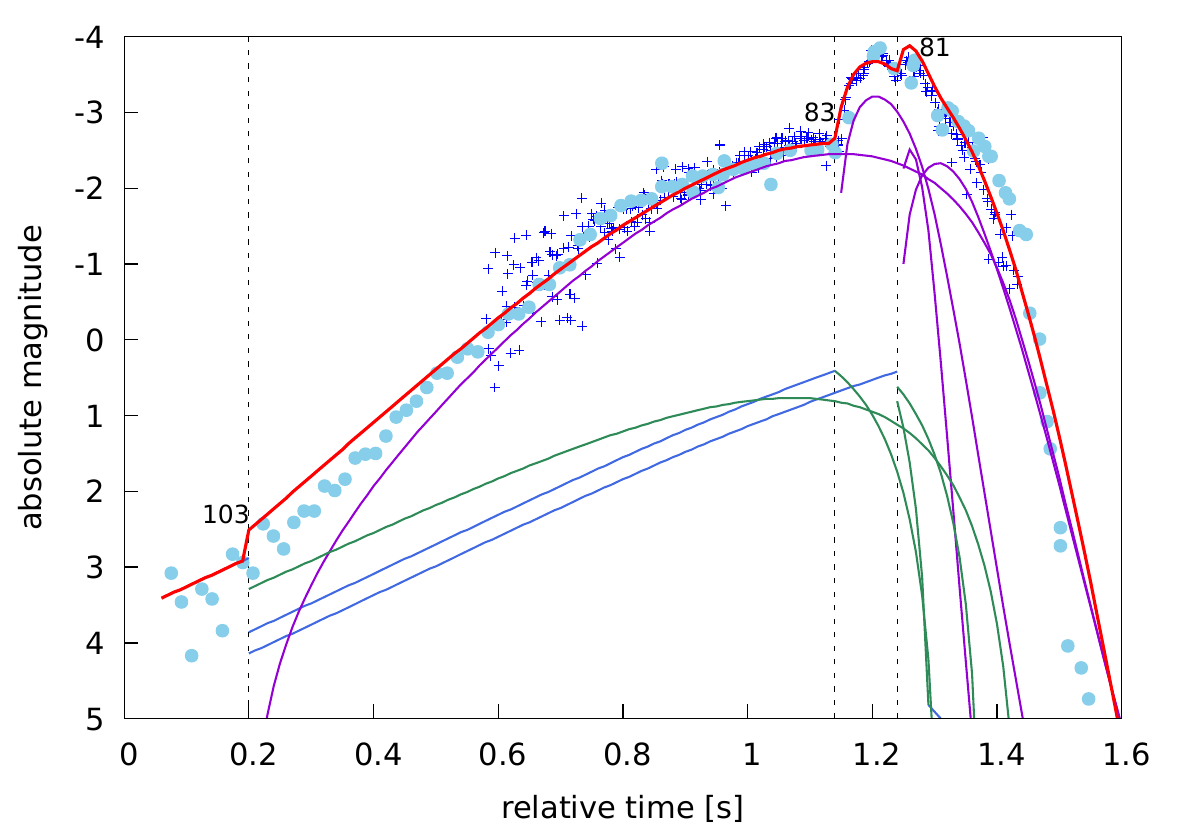}
    \caption{Geminid~10 fragmentation model is compared to the observed radiometric curve (dark-blue pluses) and a composite photometric light curve from various instruments (sky-blue disks). The total model brightness is shown as a solid red line, blue curves show the brightness of regular fragments, green curves signify eroding fragments, and violet lines indicate dust particles released from these fragments. Vertical dashed lines show fragmentation times, and numbers indicate the height above the ground in kilometers of the fragmentations.}
    \label{gem10_lc}
\end{figure}

\begin{figure}
  \centering
   \includegraphics[width=\hsize]{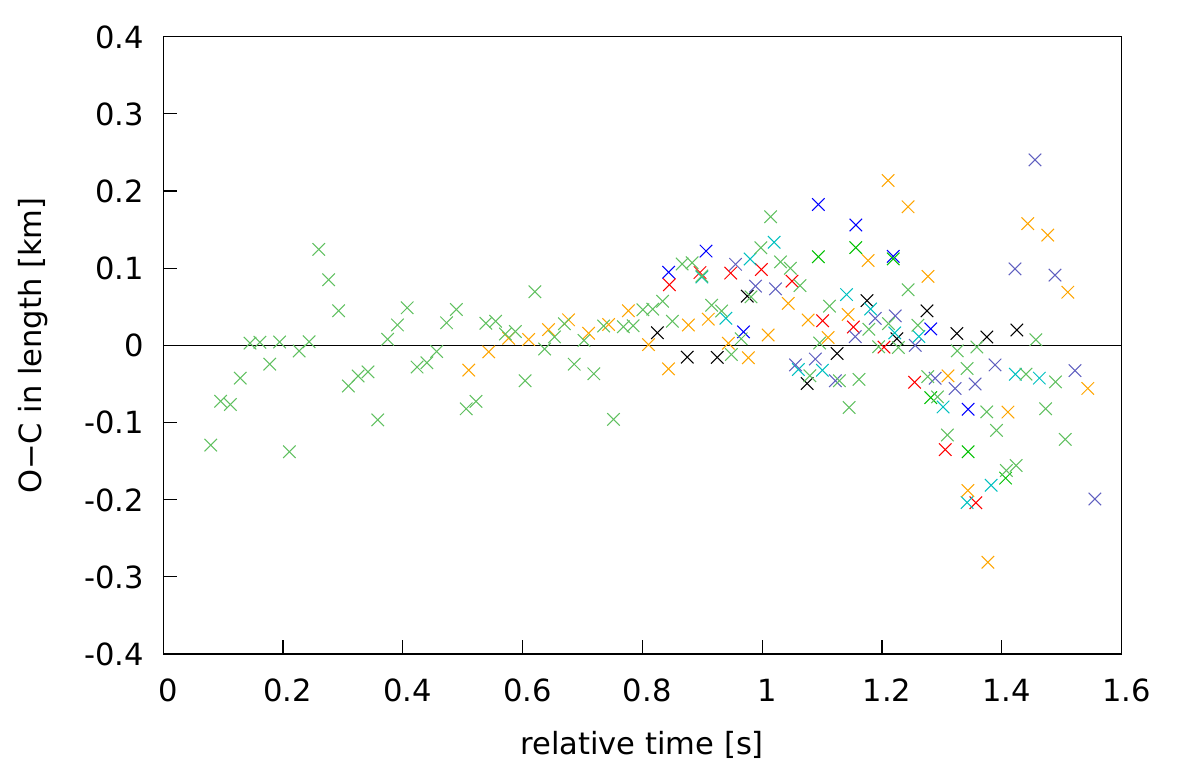}
    \caption{Geminid~10 length residuals. Different colors indicate the stations of the EN that were used to calculate the model.}
    \label{gem10_len}
\end{figure}

In some cases, a composite dataset was created from all the available sources. This enabled us to cover the widest possible range of a meteor appearances, from about $+5$ to $-4\,\rm{mag}$, without losing precision and without any detector saturation at the same time. For this reason, we were able to obtain a more robust model for these meteors that would otherwise be unreachable using any one of the observational techniques we employed. Figure~\ref{gem10_lc} shows an example of such a dataset: a composite light curve of Geminid~10 (EN131222$\_$200310). Figure~\ref{gem10_len} shows the length-along-the-trajectory residuals.

Video observations sometimes suffer from the detector saturation. However, there are empirical approaches to estimating the brightness of a meteor, even when it is saturated. In this research, we employed a novel method of calibration using radiometric observations.

The apparent magnitude of a meteor is calculated from the radiometric signal using a Pogson equation and a constant. This constant is usually derived for each EN station and each meteor from the photometrically calibrated all-sky image. However, the same constant can be used for the same radiometer and multiple meteors close in time, provided that the sky conditions do not change too much (i.e., thin clouds, the Moon). We used this method to calibrate the brightness of two Geminids captured by video cameras that were already saturated, for which photographic data were unavailable.

Some of the modeled Geminids served as the sensitivity test of the EN instruments. In particular, Geminid~20 (EN161223$\_$011634) was observed under superb conditions (a clear, moonless night with a good transparency) and the fireball appeared directly above one of the EN stations. As a result, the radiometer detected the fireball from about $-2\,\rm{mag}$, whereas under standard conditions, the detection limit of the radiometer for usable data was about $-4\,\rm{mag}$.

\subsection{Semiautomatic modeling procedure}\label{firmpik}
The modeling procedure depends on the size of the meteoroid. We used the most complex model for bright fireballs, a simplified model for moderately bright meteors and a simple erosion model for faint video meteors.

First, the fragmentation model and the optimization method for bright fireballs observed with the EN are the same as in the previous work, which focused on the fragmentation modeling of Geminid fireballs \citep{henych2024}. This model incorporates meteoroid ablation and deceleration equations, as well as two types of fragmentation: gross fragmentation, which releases several macroscopic fragments and dust grains, accompanied by a short flare; and erosion, which continuously releases dust grains, causing gradual brightening. Initially, the fragmentation times were found manually, and then optimized in the process of finding a solution. Individual fragments (and dust grains) ablate separately and the calculation stops when a fragment vanishes due to ablation or erosion or when its velocity decreases to below $2.5\,\rm{km\,s^{-1}}$. We calculated the total brightness of the meteor and the length of the foremost fragment along the trajectory and we compared them with the data. In this paper, we distinguish between fragments and grains. While fragments are usually macroscopic parts of a meteoroid at the moment of their creation, grains are smaller particles. However, there is no strict size boundary between these two categories. More importantly, they are represented differently in the modeling program.

The data optimization program, called FirMpik, is based on genetic algorithms. It starts with a population of several tens of random yet physically plausible solutions, which it then evolves. The fitness function is the inverse value of the reduced $\chi^2$ sum of the model fit to the radiometric and photometric light curves and dynamical data of the foremost fragment. We assigned empirical weights to each dataset.

We used fixed values of the grain density of $\rho_{\rm grain}=3000\,\rm{kg\,m^{-3}}$, the product of the drag ($\Gamma$) and the shape ($A$) coefficients $\Gamma A=0.8$ (continuum flow) as well as the ablation coefficient $\sigma=0.005\,\rm{kg\,MJ^{-1}}$ for Geminid fireballs. The physical model of meteoroid fragmentation used in this study was described in detail in \citet{borovicka2020b} and the semiautomatic modeling method was described in \citet{henych2023}.

Second, faint meteors are modeled using a simpler fragmentation model described in \citet{borovicka2007}. The model includes a maximum of two fragmentation times and three fragments: the original meteoroid, called the main, and two subsequent daughter fragments. One or two daughter fragments are always ablating and also eroding dust grains, while the main only ablates. The optimized model parameters are as follows: the initial mass and velocity of the meteoroid, its height, the fragmentation times, the ablation coefficient (common for all fragments and dust grains), the erosion coefficient (different for the two daughter fragments), the portion of the eroded mass from the first daughter fragment in the case of two fragmentations, and the eroded dust grain properties (where a power-law mass distribution is assumed, with maximum and minimum mass limits and a slope). In some cases, we can also constrain the meteoroid's bulk density. The fixed parameters of the model are the drag coefficient ($\Gamma=1$, free molecular flow) and the shape coefficient ($A=1.6$ for macroscopic fragments, as described in \citealt{capek2025}, and $A=1.21$ for the dust grains). The grain density is the same as that of fireballs.

In our model, we calculated a meteor profile (the brightness distribution of a meteor along its length caused by eroded dust grains) at each time step to consistently calculate the length along the trajectory of a faint meteor and also its brightness. First, we considered all dust grains that exist in a given time step. Then, we binned them by their relative length with respect to their parent fragment (basically their lag behind the fragment) and summed their light fluxes in all the bins. The bin length is optional, but should provide a higher spatial resolution of the profile than the observations. Here, we used a fixed bin length of $2\,\rm{m}$. Next, we smeared the total fluxes in the respective bins using a Gaussian profile to simulate the detection of the meteor by an optical system. This yielded the final meteor profile as simulated by the model.

We then calculated the locus of the meteor for each time step as the length of the maximum brightness in the meteor profile. The total brightness of the meteor is calculated by integrating the flux in the meteor profile from the head to the maximum length, which is again optional. We used a maximum length of $1.5\,\rm{km}$; the signal farther from the meteor's head was usually negligible. The optimizer part is the same as in \citet{henych2023}, based on parallel genetic algorithms. There are 10 free parameters for single-phase erosion and 16 for two-phase erosion.

Third, the complexity of the model for bright meteors (or faint fireballs) was between the two approaches described above. The full model of \citet{borovicka2020b} and \citet{henych2023} was used for modeling, and the locus of the meteor and its total brightness were calculated from its profile, which was calculated for each timestep. The model had a lower number of fragmentation times (and total number of fragments) than fireballs, but usually a slightly higher number than two (three), as with faint meteors. Bright flares were rarely observed in digital video camera records.

The borderline between the various modeling approaches was defined by the type of data for a given meteor because different data reduction procedures are used for different types of data. The EN data were modeled using the full model without calculating the meteor profile, while the video camera data (sometimes enhanced by the radiometer data from the EN) were modeled using either the moderately complex model or the simple erosion model from \citet{borovicka2007}. Air densities were taken from the NRLMSISE-00 atmosphere model of \citet{picone2002}.

\subsection{Manual modeling procedure}\label{manual}
Some faint videometeors (namely 32--39) were modeled both automatically and manually by one of us (VV). Our goal was to validate the semiautomatic method of modeling by comparing it to the results of the traditional method. For this, we used the same erosion model of \citet{borovicka2007}, and the model parameters were sought manually via trial and error. We then compared the quality of the fits obtained by both methods, as well as the values of the model parameters (see Sect.~\ref{meth_comp}).

\subsection{Luminous efficiency functions}\label{tau}
We used the luminous efficiency from \citet{pecina1983} for modeling faint meteors. This function depends only on meteor velocity, not its mass. For meteoroids of moderate mass ($\sim\!\!1\rm{-}40\,\rm{g}$), we tried to find a good fit with two different luminous efficiency functions: the function from \citet{pecina1983} and the function from \citet{revelle2001} modified by \citet{borovicka2020b}. We present the better of the two solutions. In this mass range, most solutions were found using the \citet{pecina1983} luminous efficiency function, but there were a few exceptions, which are indicated in Table~\ref{gem_tab1}. We used the latter function of \citet{borovicka2020b} for all meteoroids with an initial mass larger than $\sim\!40\,\rm{g}$. Considering how the luminous efficiency depends on meteoroid mass (Table~\ref{gem_tab1}), we think that there is a sharper, quantitative change in luminous efficiency for meteoroids in the $1\rm{-}40\,\rm{g}$ mass range.

\subsection{Bulk density determination}\label{rho}
To derive the bulk densities of Geminid meteoroids, we employed two methods. The first approach makes use of rather rare observations, namely, when the meteoroid is detected before the onset of erosion and the ablation is  in a steady state. From the modeling, we can derive its total mass. For fireballs, we usually assume the ablation coefficient $\sigma=0.005\,\rm{kg\,MJ^{-1}}$ and the product of $\Gamma A=0.8$. The lower the density, the larger the cross-section of the meteoroid and the brighter the fireball. For fainter meteors, we determine $\sigma$ in the optimization process, along with the bulk density, $\rho_{\rm bulk}$ (assuming $\Gamma A=1.6$). The value of the ablation coefficient also affects the meteor brightness; therefore, we must be careful when comparing densities derived from the model with widely different values of $\sigma$.

The second approach is the same as in \citet{borovicka2010}, where we fit the data with a fixed bulk density of $2000\,\rm{kg\,m^{-3}}$ and then we calculate the bulk density of the meteoroid from the erosion energy received per unit cross-section before the start of erosion and the erosion coefficient value. The empirical formula (see Eq.~2.1 in \citealt{borovicka2010}) was calibrated using observations of Draconids (low-density cometary material), often detected before the onset of erosion.

\subsection{Aerodynamic pressure}\label{dynpress}
The aerodynamic pressure acting on the meteoroid is calculated as
\begin{equation}
  p = \Gamma\rho_{\rm a} v^2,
  \label{dyn_press}
\end{equation}
where $\Gamma=1$ is the drag coefficient used to plot Geminids of all masses, $\rho_{\rm a}$ is the atmospheric density, and $v$ is the meteoroid velocity. We note that we used $\Gamma\doteq0.66$ in our fragmentation modeling of fireballs (assuming a spherical shape for the meteoroids) and $\Gamma=1$ when modeling fainter meteors. When calculating this pressure at the instant of fragmentation (seen in the radiometric or photometric data), we can simplistically assume that the mechanical strength of the bond broken during fragmentation is of the order of this aerodynamic pressure. We call this pressure a mechanical strength proxy to emphasize that it is not necessarily equal to the mechanical strength of the meteoroid or its parts. We refer to Sect.~3.1 of \cite{henych2024} for more details and a discussion of this problem.

\section{Thermal stress in the atmosphere}\label{thermal}
We estimated the magnitude of thermal stresses in Geminid meteoroids during the pre-ablation phase of atmospheric entry. We modeled the meteoroids as homogeneous spherical bodies with fast random rotation \citep{ceplecha1998}. Under these assumptions, the temperature field and the stress tensor are both spherically symmetric. However, for the calculation of these quantities, we need the dependencies of the altitude, $h$, of the meteoroid and its velocity, $v$, on time, $t$. The corresponding equations of motion are

\begin{eqnarray}
    \frac{d h}{d t} &=& -v \cos(z), \label{eqh}\\
    \frac{d v}{d t} &=& - \frac{3\Gamma}{4R} \frac{\rho_{\rm a}}{\rho_{\rm bulk}} v^2,
\end{eqnarray}
where $z$ is the zenith distance of the radiant, $\Gamma=1$, and the meteoroid radius, $R$, was assumed to be constant during the pre-ablation phase. The temperature distribution, $T(r,t)$, within a meteoroid was determined by numerical solution of the heat diffusion equation (HDE) as a function of radial distance, $r$, and time, $t$:

\begin{equation}
    \frac{\rho_{\rm bulk}\,c}{K}\,\parderi{T}{t} = \parderii{T}{r}+\frac{2}{r}\parderi{T}{r},
\end{equation}
where $c$ is the specific heat capacity and $K$ is the thermal conductivity. We imposed a zero temperature gradient at the center of the meteoroid,

\begin{equation}
     \left.\parderi{T}{r}\right|_{r=0} = 0,
\end{equation}
and based the boundary condition at the surface on energy conservation by balancing the conductive heat flux into the interior, radiative losses, and incoming energy, expressed as

\begin{equation}
    \left[K\parderi{T}{r} + \epsilon\sigma_{\rm SB} T^4\right]_{r=R} = \frac{1}{8}\Lambda \rho_{\rm a}v^3,
    \label{eq_conserv}
\end{equation}
where $\epsilon$ is emisivity of the meteoroid surface, $\sigma_{\rm SB}$ is the Stephan-Boltzmann constant, and $\Lambda$ is the energy transfer coefficient. For simplicity, this coefficient was set to $\Lambda = 1$. The atmospheric density, $\rho_{\rm a}$, is given as a function of altitude was derived from the NRLMSISE-00 model \citep{picone2002}. An initial temperature of 293\,K was assumed at an altitude of 200\,km.  We solved the HDE using the Crank--Nicolson finite difference scheme on a non-uniform radial grid, refined toward the surface, where the temperature gradients are steepest.

To determine the components of the stress tensor, we employed analytical expressions for a spherically symmetric temperature field in a solid sphere \citep{hopkinson1879}. These relations are commonly used in studies of similar problems \citep[e.g.][]{jones1966,elford1999,tirskiy2004,shestakova1997}:
\begin{eqnarray}
    \tau_{rr}(r,t) &=& \alpha \frac{2\mu (3\lambda + 2\mu)}{\lambda + 2\mu} T_{rr}(r,t),\\
    \tau_{\phi \phi}(r,t) &=& \alpha \frac{2\mu (3\lambda + 2\mu)}{\lambda + 2\mu} T_{\phi\phi}(r,t),
\end{eqnarray}
where
\begin{eqnarray}
    T_{rr}(r,t) &=& \frac{2}{R^3}\int_0^R T(x,t) x^2 dx - \frac{2}{r^3}\int_0^r T(x,t) x^2 dx, \label{eq_trr}\\
    T_{\phi\phi}(r,t) &=& \frac{2}{R^3}\int_0^R T(x,t) x^2 dx + \frac{1}{r^3}\int_0^r T(x,t) x^2 dx - T(r,t), \label{eq_tff}
\end{eqnarray}
where $\alpha$ is a linear thermal expansion coefficient, while $\lambda$ and $\mu$ are Lam\'{e} parameters. In spherical coordinates, only the components $\tau_{rr}$ and $\tau_{\phi\phi} = \tau_{\theta\theta}$ are nonzero\footnote{Here, $\phi$ and $\theta$ denote spherical coordinates, longitude and latitude.}. In the center, these components are equal, while at the surface, the radial stress vanishes (i.e., $\tau_{rr} = 0$).

The temperature profiles, $T(r,t)$, which we determined using the numerical solution of Eqs.~(\ref{eqh})--(\ref{eq_conserv}), were used to calculate the functions $T_{rr}(r,t)$ and $T_{\phi\phi}(r,t)$ according to Eqs.~(\ref{eq_trr})--(\ref{eq_tff}). The corresponding integrals were determined numerically. Using these functions, we then calculated the profiles of the stress tensor components in the meteoroid for a given altitude.

We assumed material properties representative of carbonaceous chondrites, adopting values for specific parameters mainly from \citet{capek2010}, except for the temperature dependence: bulk density $\rho_{\rm bulk} = 2260\,\mathrm{kg\,m^{-3}}$, Lam\'{e} parameters $\lambda = 17.2\,\mathrm{GPa}$ and $\mu = 17.6\,\mathrm{GPa}$, linear thermal expansion coefficient $\alpha = 8.5 \times 10^{-6}\,\mathrm{K^{-1}}$, specific heat capacity $c = 850\,\mathrm{J\,kg^{-1}\,K^{-1}}$, thermal conductivity $K=1\,\mathrm{W\,m^{-1}\,K^{-1}}$, and tensile strength $\sigma_{\rm t} = -10\,\mathrm{MPa}$.

\section{Results}\label{results}
In this section, we present the results of the fragmentation modeling of the observed Geminid fireballs and faint meteors, as well as the results of the numerical modeling of the thermal stress exerted on the model meteoroids. The observed Geminids are characterized in Table~\ref{gem_tab1}.

\begin{table*}
\caption{Geminid meteors that were modeled in this study and in \citet{henych2024}.}
\label{gem_tab1}
\centering
\begin{tabular}{c c S[table-format=2.1] c S[table-format=2.1] c S[table-format=2.1] c}
\hline\hline
no. & fireball/meteor & {mass} & velocity & {diameter} & trajectory & {maximum} & luminous efficiency\\
 & code & {[kg]} & [km/s] & {[cm]} & slope [$^\circ$] & {magnitude} & function\\
\hline
1 & EN151220$\_$024219 & \num{0.83E+0} & 35.21 & 8.6 & 63.0 & -11.9 & 1\\
2 & EN141221$\_$194403 & \num{0.13E+0} & 36.15 & 4.6 & 35.6 & -10.1 & 1\\
3 & EN131221$\_$024216 & \num{2.7E-2} & 35.69 & 2.7 & 63.4 & -8.0 & 1\\
4 & EN141220$\_$185349 & \num{2.4E-2} & 36.09 & 2.6 & 25.4 & -7.9 & 1\\
5 & EN121218$\_$193710 & \num{1.7E+0} & 35.60 & 10.8 & 29.2 & -11.5 & 1\\
6 & EN131218$\_$012640 & \num{0.13E+0} & 35.03 & 4.6 & 71.8 & -9.6 & 1\\
7 & EN091220$\_$210049 & \num{3.3E-2} & 35.53 & 2.9 & 44.7 & -8.0 & 1\\
8 & EN141219$\_$200732 & \num{0.16E+0} & 35.61 & 4.9 & 37.3 & -9.3 & 1\\
9 & EN141222$\_$220653 & \num{0.27E+0} & 36.12 & 6.0 & 53.0 & -10.7 & 1\\
10 & EN131222$\_$200310 & \num{2.0E-3} & 35.77 & 1.2 & 34.7 & -3.8 & 2\\
11 & EN131222$\_$210107 & \num{9.4E-3} & 35.73 & 1.9 & 42.3 & -5.6 & 2\\
12 & EN141222$\_$221649 & \num{1.6E-2} & 35.48 & 2.3 & 53.3 & -5.5 & 2\\
13 & EN141222$\_$232812 & \num{3.5E-2} & 35.88 & 3.0 & 65.1 & -6.8 & 2\\
14 & EN151222$\_$004757 & \num{3.8E-2} & 35.78 & 3.1 & 71.4 & -7.0 & 2\\
15 & EN131222$\_$173127 & \num{5.1E-4} & 36.21 & 0.7 & 13.0 & -1.3 & 2\\
16 & EN131222$\_$174526 & \num{5.8E-4} & 36.46 & 0.8 & 15.0 & -1.7 & 2\\
17 & EN141221$\_$205216 & \num{6.5E-2} & 36.00 & 3.7 & 46.7 & -9.0 & 1\\
18 & EN141223$\_$182231 & \num{0.14E+0} & 35.96 & 4.8 & 21.7 & -11.7 & 1\\
19 & EN141221$\_$234415 & \num{1.4E-2} & 35.85 & 2.2 & 70.6 & -8.1 & 1\\
20 & EN161223$\_$011634 & \num{8.2E-3} & 35.89 & 1.8 & 71.2 & -5.4 & 2\\
21 & 22c13123 & \num{1.3E-4} & 36.12 & 0.5 & 24.8 & -0.2 & 2\\
22 & 22c13127 & \num{5.6E-5} & 36.63 & 0.3 & 27.7 & -1.7 & 2\\
23 & 22c13157 & \num{4.3E-4} & 36.41 & 0.7 & 38.7 & -2.4 & 2\\
24 & EN071221$\_$223236 & \num{1.2E-2} & 35.61 & 2.1 & 62.2 & -4.9 & 2\\
25 & EN121221$\_$002615 & \num{1.7E-2} & 36.38 & 2.4 & 75.1 & -5.7 & 2\\
26 & EN121221$\_$052200 & \num{6.8E-3} & 35.75 & 1.7 & 40.0 & -4.7 & 2\\
27 & EN141221$\_$042454 & \num{3.8E-3} & 35.94 & 1.4 & 46.6 & -4.2 & 2\\
28 & EN141221$\_$044702 & \num{1.8E-2} & 35.65 & 2.4 & 42.4 & -5.3 & 2\\
29 & 22c13146 & \num{2.7E-4} & 36.40 & 0.6 & 36.3 & -1.2 & 2\\
30 & 22c13164 & \num{4.8E-5} & 36.47 & 0.3 & 42.2 & -0.3 & 2\\
31 & 22c13196 & \num{2.1E-5} & 36.69 & 0.3 & 59.8 & 0.7 & 2\\
32 & 22c12237 & \num{1.1E-5} & 36.25 & 0.2 & 59.2 & 1.3 & 2\\
33 & 22c12256 & \num{5.7E-6} & 37.16 & 0.2 & 53.3 & 1.7 & 2\\
34 & 22c12270 & \num{3.0E-5} & 36.78 & 0.3 & 49.2 & 0.2 & 2\\
35 & 22c12272 & \num{1.5E-5} & 35.78 & 0.2 & 48.9 & 0.7 & 2\\
36 & 22c12276 & \num{1.1E-5} & 35.27 & 0.2 & 47.3 & 1.7 & 2\\
37 & 22c12277 & \num{1.3E-5} & 35.78 & 0.2 & 46.9 & 1.6 & 2\\
38 & 22c12283 & \num{1.5E-5} & 36.30 & 0.2 & 43.6 & 1.0 & 2\\
39 & 22c12297 & \num{4.3E-5} & 36.23 & 0.3 & 40.5 & 0.6 & 2\\
\hline
\end{tabular}
\tablefoot{The table shows their initial mass, entry velocity, estimated diameter ($\rho_{\rm bulk}=2500\,\rm{kg\,m^{-3}}$), trajectory slope to the ground, maximum magnitude, and luminous efficiency function used in modeling (1=\citet{borovicka2020b}, 2=\citealt{pecina1983}).}
\end{table*}

\subsection{Aerodynamic pressure statistics}
\begin{figure}
  \centering
   \includegraphics[width=\hsize]{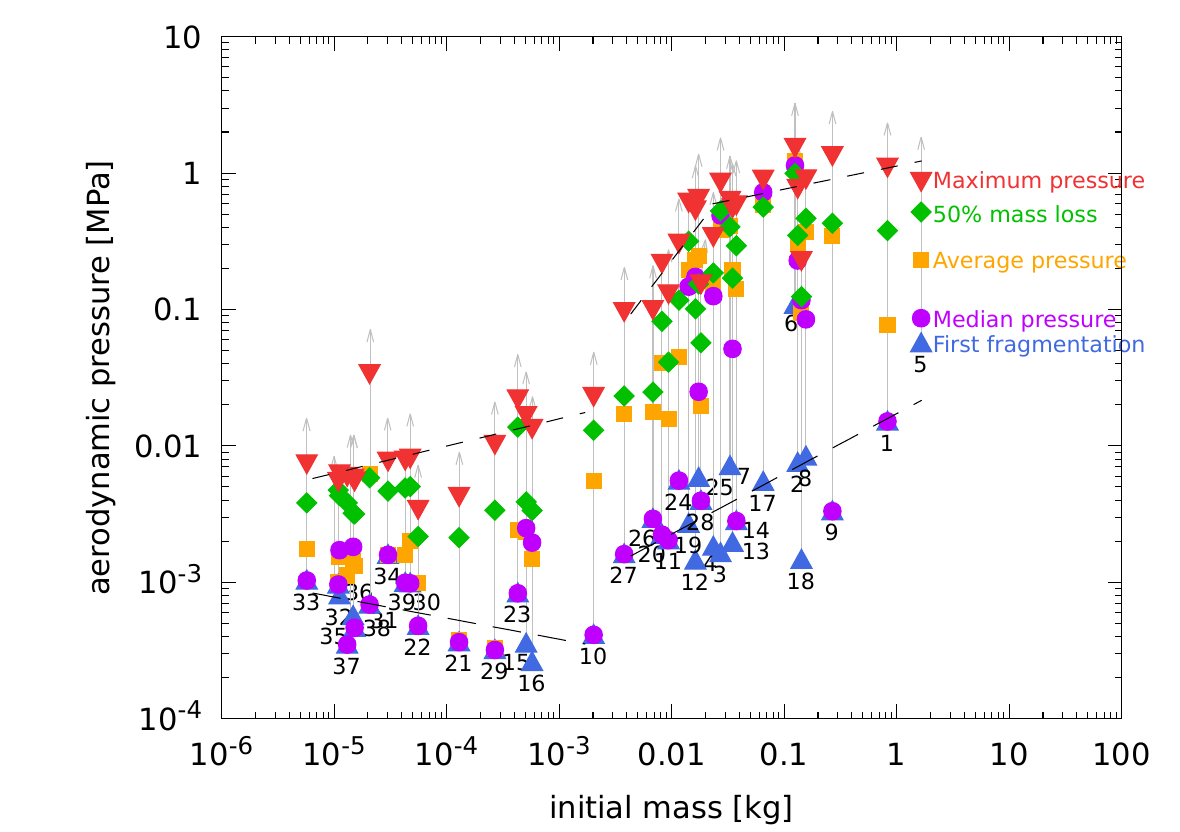}
    \caption{Statistical values of aerodynamic pressure for Geminids. Blue triangles show the pressure for the first observed fragmentation, green diamonds, magenta disks, and orange squares show the average pressure in the volume of meteoroids, calculated in different ways, and red triangles show the maximum attained pressure. The gray arrow indicates that the maximum pressure is only a lower limit. The numbers in the plot are meteor designations from Table~\ref{gem_tab1}. The dashed lines are power-law fits to minimum or maximum dynamic pressures. Both axes are logarithmic.}
    \label{strength_stat}
\end{figure}

The plot in Fig.~\ref{strength_stat} shows an overall aerodynamic pressure statistics: the initial fragmentation pressure, the pressure at which $50\%$ of the initial mass was lost, the median pressure and the weighted average pressure within the meteoroid volume and the maximum pressure exerted on any part of the meteoroid \citep{henych2024}. The latter is usually the lower limit, because a small part typically survives the final fragmentation, and the maximum pressure already does not cause its fragmentation; therefore, it must be stronger.

For small Geminids with initial masses below $2\,\rm{g}$, we observed only marginal trends in the minimum, average, and maximum aerodynamic pressure. The values exhibit a rather large variance, with a minimum pressure between $0.3$ and $1.6\,\rm{kPa}$ and a maximum pressure between $3.4$ and $34\,\rm{kPa}$. The median pressure is usually the same as the minimum pressure, although this is not always the case.

For Geminids with moderate masses ($2\,\rm{g} < {\it m} < 20\,\rm{g}$), we observe the steepest change in the maximum aerodynamic pressure going from a few tens of kPa to several hundreds of kPa. For even larger Geminids, the maximum pressure plateaus at around $1\,\rm{MPa}$. In both these groups of Geminids, the minimum fragmentation pressure increases with an increasing meteoroid mass.

\subsection{Mass loss modes}
\begin{figure}
  \centering
   \includegraphics[width=\hsize]{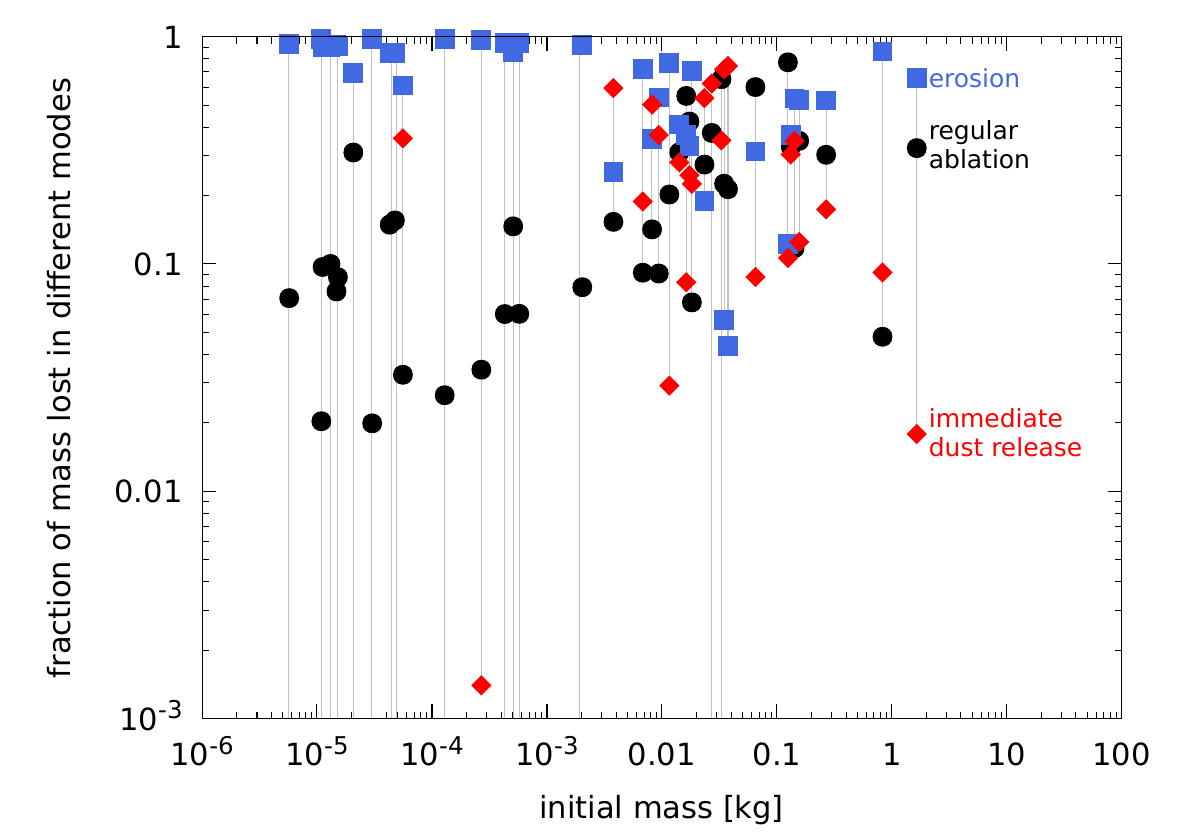}
    \caption{Regimes of mass loss vs. initial meteoroid mass. Both axes are logarithmic.}
    \label{maslost}
\end{figure}

The plot in Fig.~\ref{maslost} describes exactly how the Geminids lose mass in the atmosphere. Our model includes three processes: regular ablation, immediate dust release after a gross fragmentation, and gradual erosion of dust grains. For faint video meteors, we typically do not observe any immediate dust release; the majority of the mass is lost via erosion. The only important exception is Geminid 22, which we discuss further below in Sect.~\ref{discussion}. The importance of erosion in Geminids depends on mass in a systematic way and the same is true for regular ablation.

\subsection{Strength distribution}
\begin{figure}
  \centering
   \includegraphics[width=\hsize]{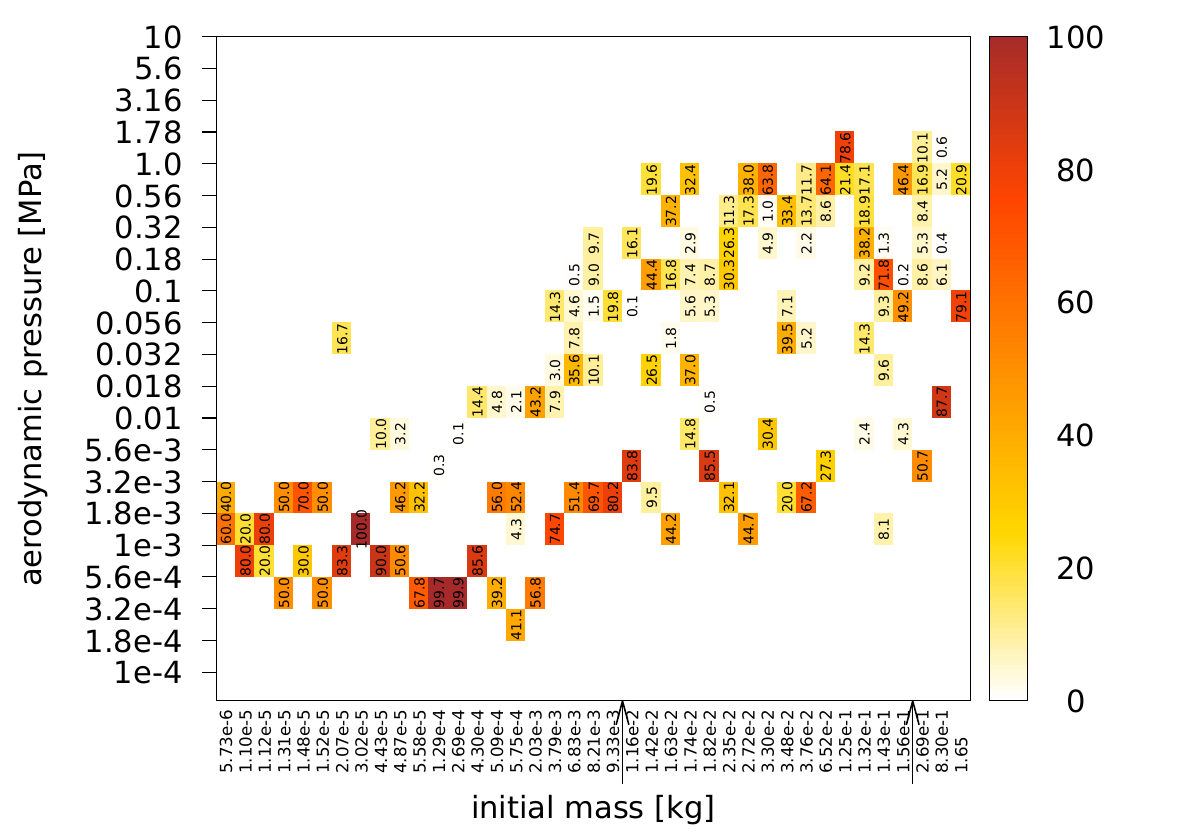}
    \caption{Aerodynamic pressure distribution vs. the initial mass of the meteoroid for modeled Geminids. The predefined bins span five orders of magnitude to describe meteors of all sizes. Numbers and colors show the percentage of the entry mass destroyed at a given aerodynamic pressure. Arrows indicate the borderlines between distinct groups of Geminids; see the text for details.}
    \label{strength_dist}
\end{figure}

The strength proxy distribution in the meteoroid is shown in Fig.~\ref{strength_dist}. We calculated it using the procedure described in \citealt{henych2024} (see their Sect.~3.1 and 4.2). The figure shows the proportion of the original mass that the Geminids lose at a given aerodynamic pressure through gross fragmentation or erosion (for which we indicate the pressure of its start). Here, we observe two distinct groups. The first group includes smaller meteoroids ($m\lesssim0.012\,\rm{kg}$) and the largest meteoroids ($m\gtrsim0.26\,\rm{kg}$), which lose most of their initial mass in the first fragmentation. However, there are exceptions to this rule, and the value of the pressure, causing the initial fragmentation, changes by two orders of magnitude. The second group comprises meteoroids of moderate mass that lose mass more gradually. We also see similar trends and structures as in Fig.~\ref{strength_stat}, but they are slightly smeared by the binning of the pressure values in this plot.

\subsection{Mass evolution}
\begin{figure}
  \centering
   \includegraphics[width=\hsize]{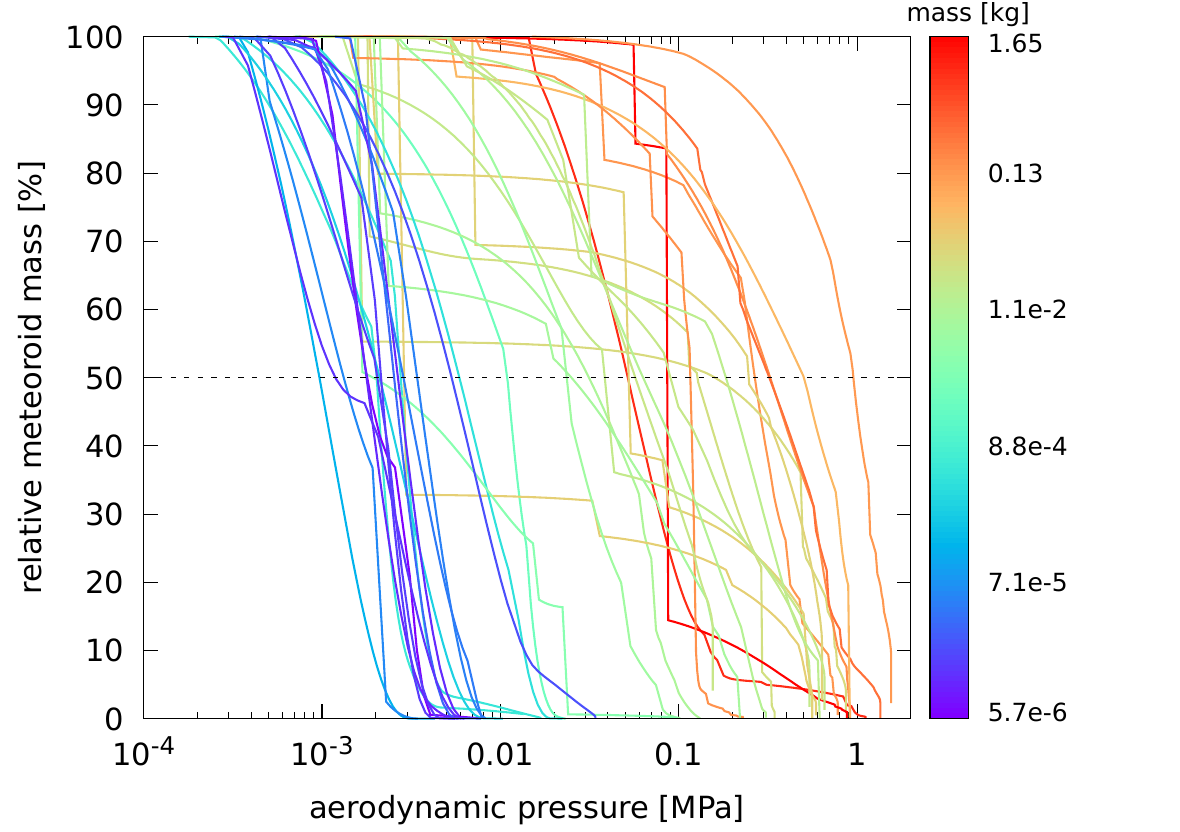}
    \caption{Relative meteoroid mass evolution as a function of the dynamic pressure (logarithmic scale) and the initial mass (color coded, also logarithmic scale).}
    \label{mass_evo}
\end{figure}

Figure~\ref{mass_evo} shows how the total mass of all macroscopic fragments (divided by the initial mass of meteoroids) changes with the aerodynamic pressure. The individual Geminids are color-coded according to their initial mass. As we can see, both ends of the mass spectrum (blue and orange to red) show a more gradual change in mass caused by dust erosion, while the meteoroids in between (cyan to green) exhibit more abrupt changes in mass.

\subsection{Grain sizes}
\begin{figure}
  \centering
   \includegraphics[width=\hsize]{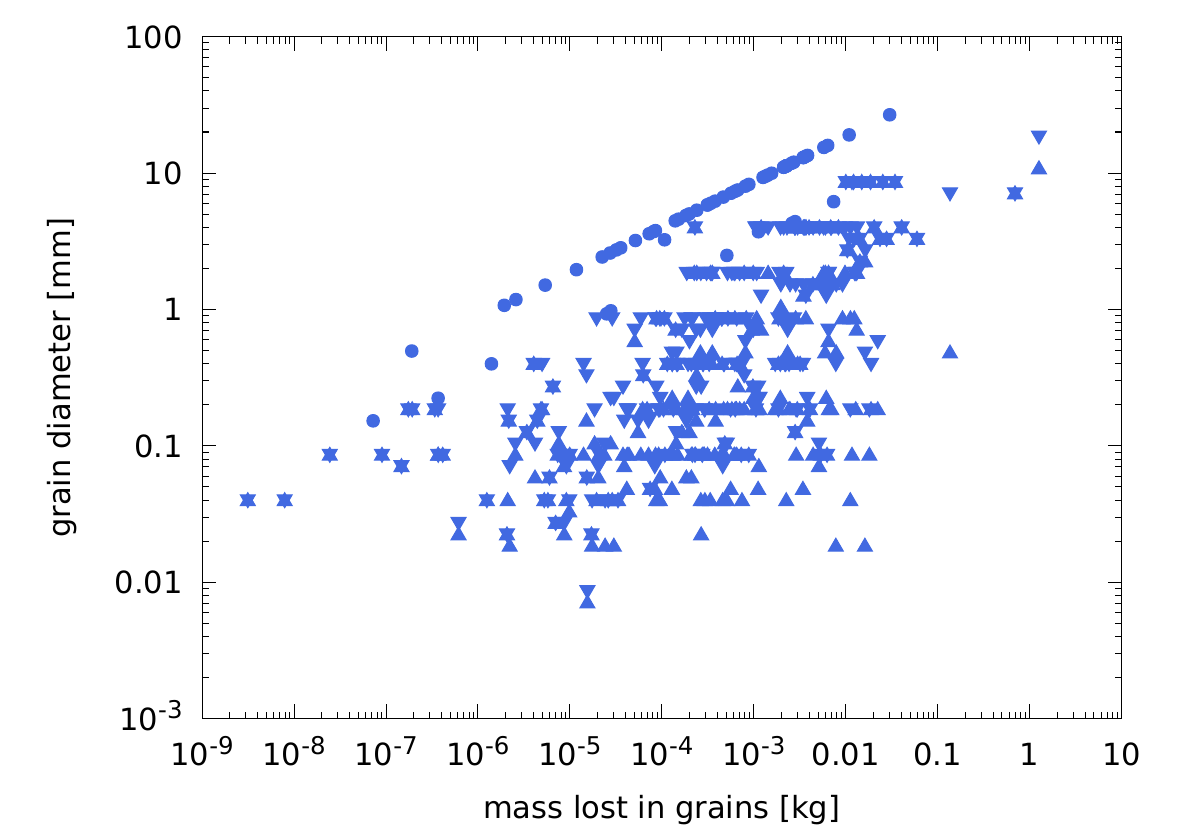}
    \caption{Grain sizes of eroded mass and dust released in gross fragmentation vs. mass lost in such events. Lower grain sizes are shown with $\blacktriangle$, upper grain sizes are shown with $\blacktriangledown$. The disks show the sizes of the non-fragmenting parts and lie on a line with a slope of 1/3. Some of the disks lie below this line due to the multiplicity of these fragments (see the text for details). Both axes are logarithmic.}
    \label{grains}
\end{figure}

One of the model parameters is the mass range of grains released either through erosion or in gross fragmentation as dust. We calculated the sizes of these grains assuming shperical shapes and a grain density of $\rho_{\rm grain}=3000\,\rm{kg\,m^{-3}}$. These grains are plotted in Fig.~\ref{grains} as a function of the mass released in any fragmentation event. This means that there are several points in the plot for a single Geminid. Moreover, two points are plotted for each fragmentation: a lower grain size ($\blacktriangle$) and an upper grain size ($\blacktriangledown$). When these two sizes are equal, the two symbols merge into a star symbol. We also plot the non-fragmenting parts of meteoroids as blue disks.

The grain sizes cover the plot without substantial grouping, mostly ranging from $20\,\rm{\mu m}$ to nearly 2 centimeters. The smallest grain size observed is $7\,\rm{\mu m}$, but it was only observed in a single Geminid. The rows of grain sizes in the plot are caused by the discrete values of the grain mass limits used in the model (four values per decade of mass). The non-fragmenting parts lie on a line with a slope of 1/3. Some of them lie below this line. These are multiple identical fragments released in a single fragmentation and we plot size of a single fragment versus their total mass.

\subsection{Mass-velocity plot}
\begin{figure}
  \centering
   \includegraphics[width=\hsize]{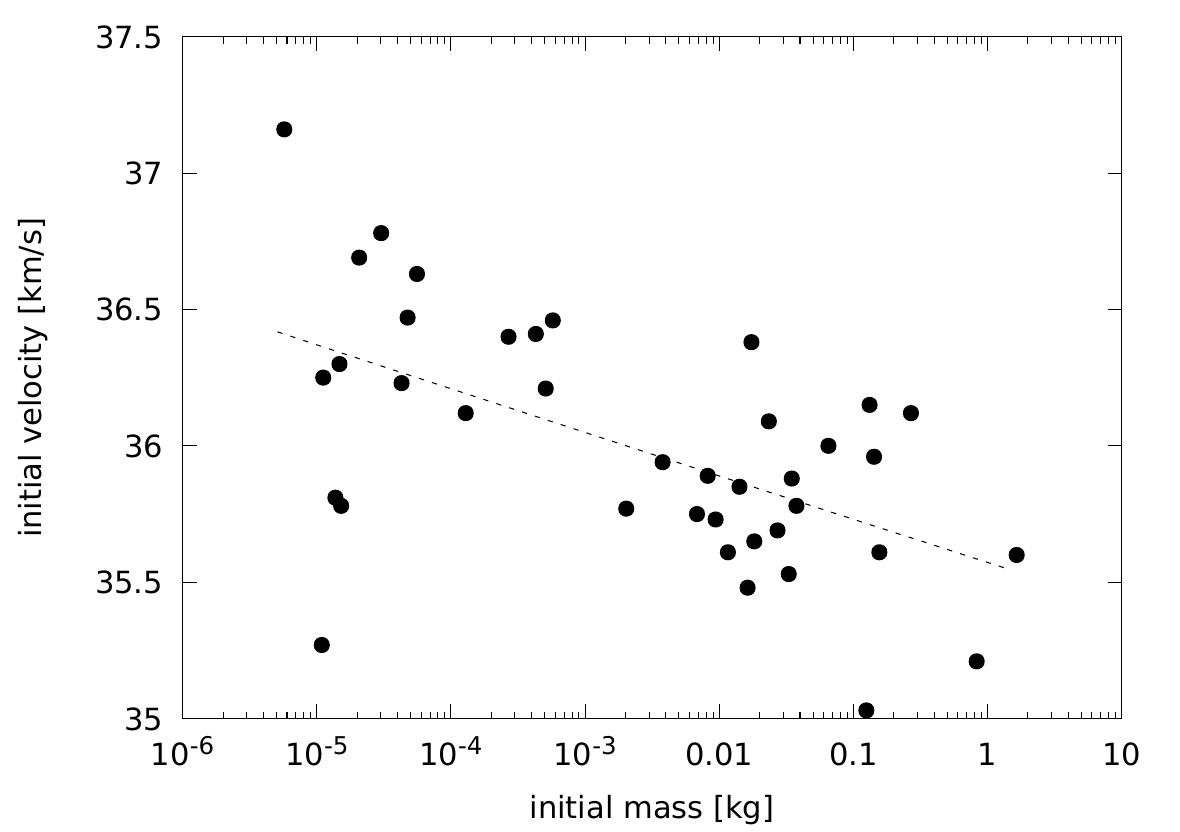}
    \caption{Initial velocity vs. initial meteoroid mass both derived from the model.}
    \label{mass_velocity}
\end{figure}

From the model, we can also derive a more precise estimate of the initial velocity and mass of the meteoroid. When plotted in Fig.~\ref{mass_velocity}, we can see that they are not randomly distributed. The initial velocity decreases with an increasing mass.

\subsection{Thermal stress modeling}
We estimated the evolution of temperature and thermal stress of each modeled meteoroid based on its initial mass, radiant zenith angle, and entry velocity. Subsequently, we determined the failure height above the ground at which the Griffith criterion for brittle failure is first satisfied in any part of the meteoroid's volume. At this height, we also calculated the aerodynamic pressure exerted on the front part of the meteoroid so that we could compare it directly with pressures derived from fragmentation modeling of Geminids.

Our model reproduces the expected and well-known result that only a thin surface shell of the meteoroid is significantly heated. For the assumed thermophysical parameters, the thickness of this heated layer is around $1\,\rm{mm}$, while the interior remains at its original temperature. Significant heating of the interior occurs only in the case of the smallest bodies, with a radius comparable to the thickness of this thermal skin depth. Regarding thermal stresses, the surface experiences compressive stress, while the interior experiences tensile stress. In all modeled cases, the material strength is exceeded before the surface temperature reaches the melting point ($\sim$1800\,\rm{K}). The model, designed to describe the pre-ablation phase, is therefore consistent in this respect.

\begin{figure}
  \centering
   \includegraphics[width=\hsize]{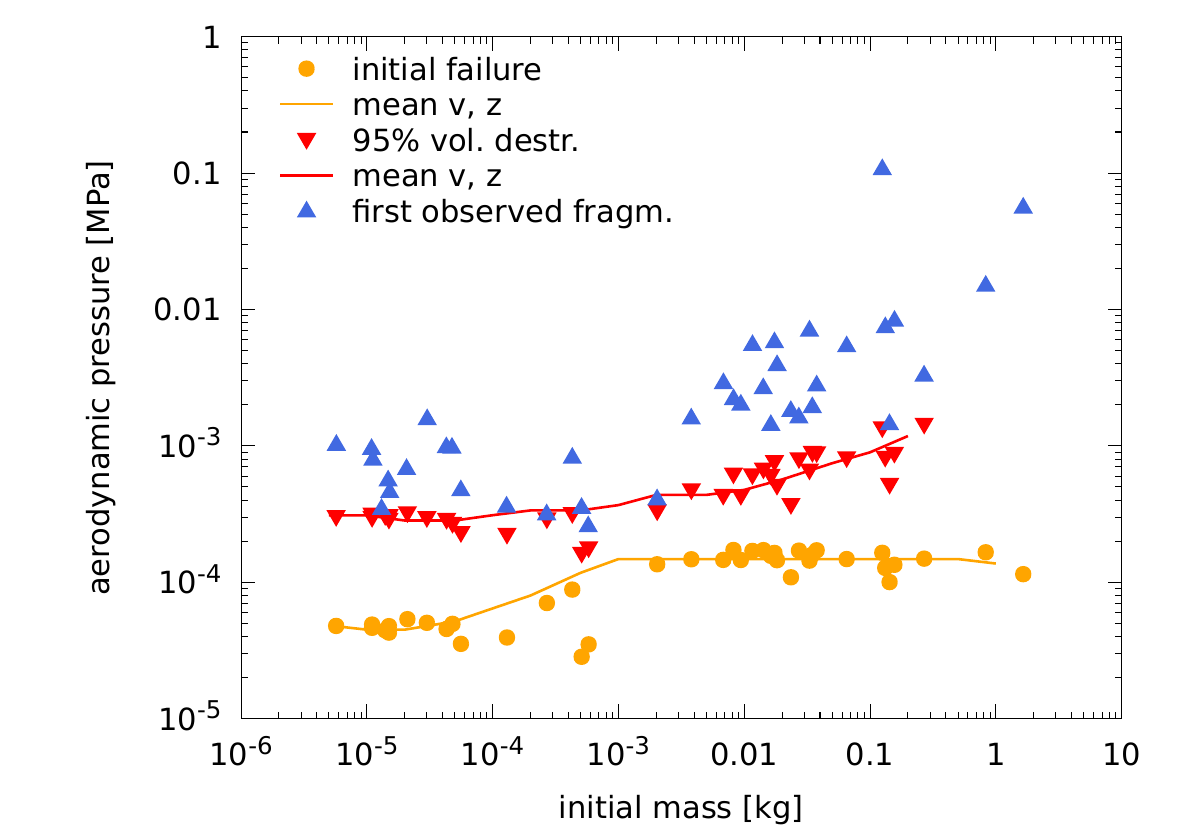}
\caption{Aerodynamic pressures at which thermal stress exceeds the material strength somewhere in the meteoroid volume (orange disks), in $95\%$ of the meteoroid volume (red triangles), and the pressure at which the first fragmentation was observed (blue triangles). The solid curves show the fragmentation in meteoroid models with an average initial velocity ($36.0\,\rm{km\,s^{-1}}$) and a radiant zenith distance ($43.5^\circ$).}
\label{stress}
\end{figure}

In Fig.~\ref{stress}, we compare aerodynamic pressures for the initial meteoroid fragmentation (blue triangles; see also Fig.~\ref{strength_stat}) and the outcomes of the thermal stress modeling (other symbols and solid lines). Both the observed data and the thermal stress simulation results reveal two distinct aerodynamic pressure values of the initial fragmentation associated with different initial meteoroid mass ranges and with a smooth transition between $0.1\,\rm{g}$ and $1\,\rm{g}$.

\begin{figure}
  \centering
   \includegraphics[width=\hsize]{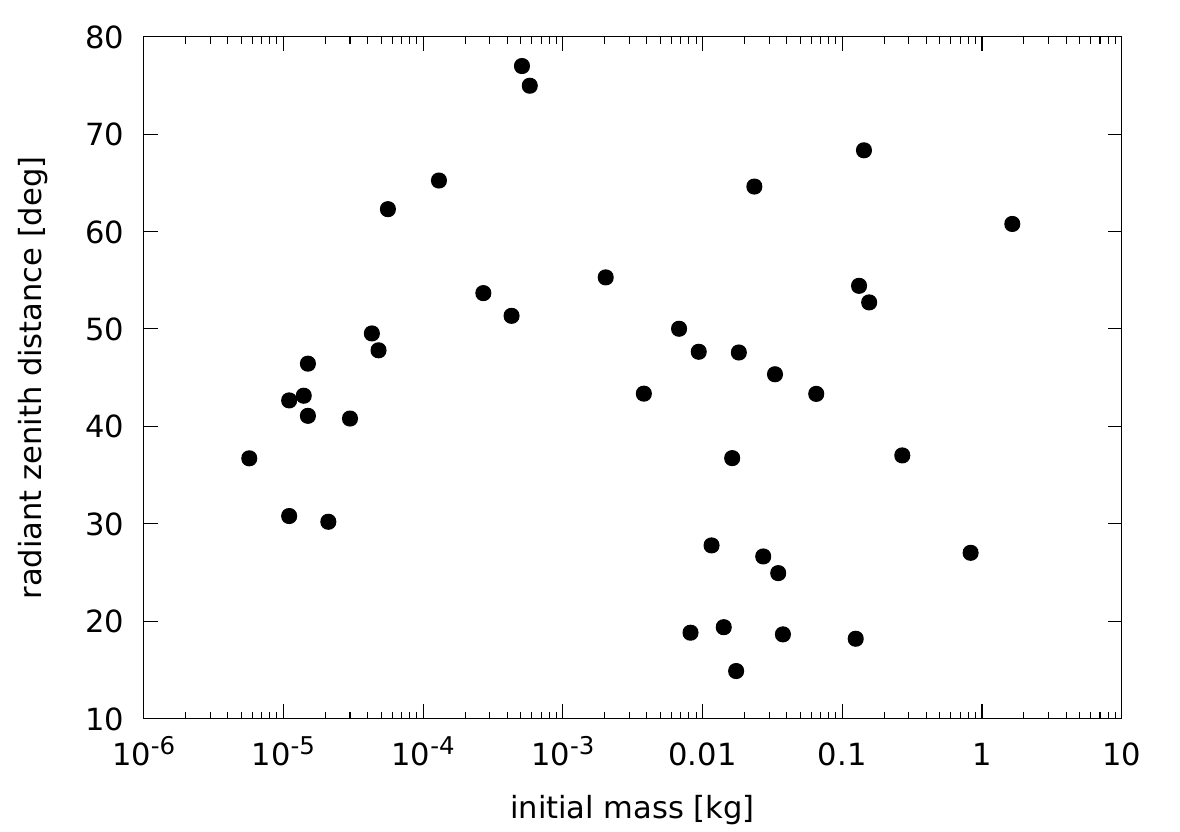}
    \caption{Zenith distance of the radiants of the observed Geminids vs. initial meteoroid mass.}
    \label{mass_zd}
\end{figure}

The dependence of the failure aerodynamic pressure on meteoroid mass is partly obscured by the fact that individual meteors have different zenith angles of radiant (and, to a lesser extent the initial velocity), which introduces a noticeable scatter into the results. We show the distribution of the zenith distance of the radiants of the observed Geminids vs. the initial meteoroid mass in Fig.~\ref{mass_zd}. While the radiants of the Geminids with masses larger than $\sim\!\!1\,\rm{g}$ are randomly distributed, the radiants of smaller Geminids are not.

Despite the scatter, two distinct altitude regimes can be identified. For meteoroids with masses below approximately $0.1\,\rm{g}$, the predicted initial failure aerodynamic pressure is around $50\,\rm{Pa}$ (an altitude of around $116\,\rm{km}$). For more massive meteoroids, the corresponding aerodynamic pressure is around $150\,\rm{Pa}$ (an altitude of around $109\,\rm{km}$). We can compare these values to the average observed beginning heights of the modeled Geminids. For Geminids with an initial mass smaller than $0.1\,\rm{g}$ it is $101\pm2\,\rm{km}$; for more massive Geminids, it is $99\pm4\,\rm{km}$.

We must emphasize one more difference between smaller and larger meteoroids. In Fig.~\ref{stress}, we plot the aerodynamic pressure assuming $\Gamma=1$ (Eq.~\ref{dyn_press}) for all the Geminids in this study, but we note that in this specific case, this assumption is a bit misleading. This value is appropriate for small meteoroids traveling in free molecular flow; however, for large meteoroids penetrating deeper into the atmosphere (continuum flow), the actual value of the drag coefficient is $\Gamma\doteq0.66$, assuming a spherical shape. This would decrease the pressure values accordingly.

Our simple model does not aim to reproduce the observed fragmentation aerodynamic pressures precisely, as we cannot expect fragmentation to occur exactly at the altitude at which the Griffith criterion for brittle failure is first satisfied somewhere within the volume of the meteoroid. Nevertheless, the model captures a relatively clear separation in the initial failure pressures for meteoroids above and below a certain mass threshold, as well as a trend of failure in $95\%$ of the meteoroid volume. This is in line with the behavior derived from the modeling of the fragmentation of the Geminids.

\subsection{Method comparison}\label{meth_comp}
\begin{figure}
  \centering
   \includegraphics[width=\hsize]{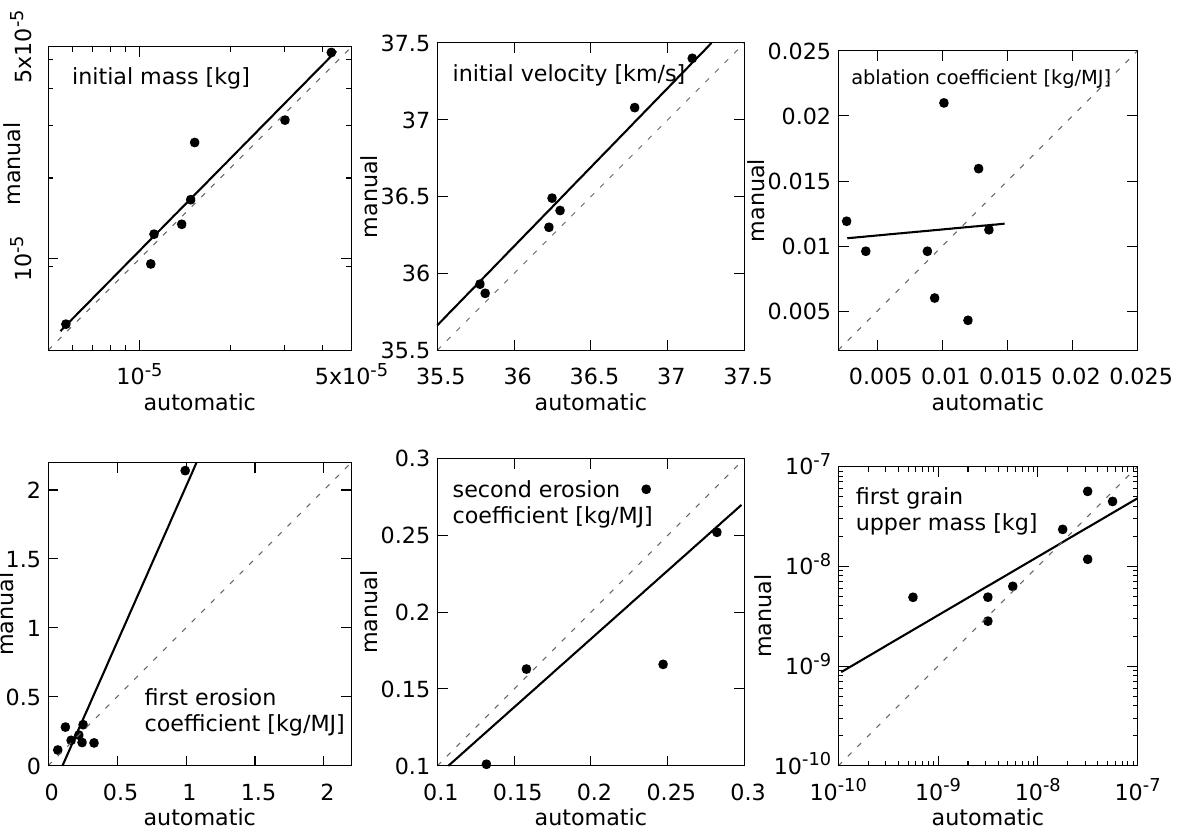}
\caption{Comparison of the semiautomatic and manual methods of modeling for six model parameters: initial mass, initial velocity, ablation coefficient, erosion coefficients for the first and second erosion (if there was any), and grain upper mass limit for the first erosion. The parameters were fit by a linear or power-law function, a dashed line indicates direct proportionality.}
\label{model_comp}
\end{figure}

Figure~\ref{model_comp} shows a comparison of the semiautomatic and manual modeling methods for six model parameters. Both methods consistently derived the initial mass and initial velocity of the meteoroids. The ablation coefficient varied moderately, and the ratio of the values from the two methods was greater than two in some cases. The erosion coefficient for the first erosion was derived similarly by both methods, except for one meteor. The erosion coefficients for the second erosion derived by both methods were similar (within a factor of 1.6), as was the upper grain mass limit (within a factor of two).

Overall, the quality of the fits of the data for both the semiautomatic and manual methods was similar, but the residual sum was smaller for the manual solution for all but one meteor. It appears that the manual approach can yield slightly better fits of the data. On the other hand, the semiautomatic method is usually faster, because it can deliver a good solution in a couple of hours on a legacy computer cluster with 48 threads (24 physical cores).

\subsection{Bulk density}
We were only able to derive the bulk density for three Geminids using the first approach (Sect.~\ref{rho}). All of the meteoroids were small and had been clearly observed before the first fragmentation (Geminids 10, 30, 39). The results are given in Table~\ref{gem_tab2}. We assumed a grain density of $\rho_{\rm grain}=3000\,\rm{kg\,m^{-3}}$ and the average ablation coefficient derived for these meteors from the model was $0.011\pm0.002\,\rm{kg\,MJ^{-1}}$. The uncertainty of the derived densities was determined to likely be on the order of $400\,\rm{kg\,m^{-3}}$ based on the spread of the bulk density values derived from several good fits.

We used the second approach for 8 faint Geminids, which we modeled both automatically and manually (32--39) and the resulting bulk densities are in the second part of Table~\ref{gem_tab2}. The uncertainty is probably of the same order as in the first approach. We note that there is a rather large discrepancy in the derived bulk densities for Geminid~39 between the two methods.

\begin{table}
\caption{Bulk densities of Geminids derived from the modeling.}
\label{gem_tab2}
\centering
\begin{tabular}{c c}
\hline\hline
no. & density [$\rm{kg/m^3}$]\\
\hline
10 & 3000\\
30 & 1800\\
39 & 2100\\
\hline
32 & 2800\\
33 & 2400\\
34 & 2700\\
35 & 2100\\
36 & 2500\\
37 & 1400\\
38 & 2500\\
39 & 2800\\
\end{tabular}
\tablefoot{The top panel shows bulk densities derived from pre-erosion observations, while the bottom panel shows bulk densities calculated from the erosion energy. The results are rounded to the hundreds and the uncertainties are on the order of $400\,\rm{kg\,m^{-3}}$.}
\end{table}

\section{Discussion}\label{discussion}
Here, we start with a comparison of the results derived from observations of Geminids and from numerical thermal stress simulations. The sudden increase in minimum fragmentation pressure observed in the Geminids (see Fig.~\ref{strength_stat}) is correlated with the smooth transition of failure due to thermal stress (see Fig.~\ref{stress}). This suggests that thermal stress drives this change and it could be an important process eroding meteoroids. Figure~\ref{stress} shows that thermal stress can disrupt the entire meteoroid before the aerodynamic pressure needed for initial observed fragmentation is reached. Mechanical forces then probably only complete the work of destruction. We note that for the largest meteoroids in our sample, the approximation of fast random rotation may fail, and their behavior may differ.

The thickness of the layer crushed by thermal stress is similar for both small and large meteoroids. It is relatively much larger for small meteoroids, which can be completely disrupted before they begin to ablate and shine. Lower in the atmosphere, they begin to erode dust grains or their conglomerates. Their size probably depends on the strength of the thermal stress, which, in turn, depends on the material properties of the meteoroid (thermal conductivity and mechanical strength).

In larger meteoroids, the surface layer is thin relative to their overall size. Thermal stress damages the surface layer, which explains the initial erosion seen in many Geminids. Further fragmentation processes deeper in the atmosphere then reveal the inner structure of the meteoroids. For the largest ones ($m>20\,\rm{g}$), we can probably derive a mechanical strength proxy (see Sect.~\ref{dynpress}) of the most compact and coherent parts that characterize the carbonaceous material itself.

Our numerical model can only detect where the thermal stress in the model body exceeds the Griffith criterion. It cannot predict what happens next. It is conceivable that one or more fractures have been created; however, it is also possible that a network of fractures formed, decreasing the overall strength of the meteoroid. The formation of a fracture substantially affects the thermal stress in the body. We assume that the load is released to some degree.

Now, we need to explain the sudden change in the observed initial fragmentation pressure and in the numerical simulations of failure due to thermal stress in Fig.~\ref{stress} for model meteoroids with the same initial mass, entry velocity, and radiant zenith distance as the observed Geminids. To the contrary, when the failure criterion was calculated using an average entry velocity ($36.0\,\rm{km\,s^{-1}}$) and a radiant zenith distance ($43.5^\circ$), the transition from small to large meteoroids was smooth (solid curves in Fig.~\ref{stress}).

This is probably because the radiant zenith distance of the faint Geminids in our sample is not randomly distributed but rather shows a trend (see Fig.~\ref{mass_zd}). Fainter meteors were observed using cameras with a narrow or medium field of view (see Sect.~\ref{data_mod}), which could have introduced an observational bias to this sample. Another potential source of bias is the selection of meteors for modeling and poor weather conditions at one of the observatories for part of the night, resulting in a lack of two-station meteors for certain zenith distance ranges. We note that the faint Geminids (10, 15, 16, 21--23, and 29--39) were observed on two nights around the maximum activity of the Geminids on December 13, 2022.

The aerodynamic pressures of the initial fragmentation in Fig.~\ref{stress} are more dispersed than the values derived from the thermal stress modeling. This can be explained by the varying shapes (different actual value of $\Gamma A$), mechanical strengths, and porosities of different meteoroids, as well as their stochastic failure behavior.

Figure~\ref{maslost} shows that gradual erosion of dust grains (blue squares) dominates in small meteors, then drops to zero at around $30\,\rm{g}$ (Geminids 3 and 7),  rising back to tens of percent for larger Geminids. The importance of a regular ablation (black disks) gradually increases with initial mass and peaks between $\sim\!\!\!20\rm{-}200\,\rm{g}$, then it decreases again (noting that we only modeled three fireballs with $m>200\,\rm{g}$). An immediate dust release (red diamonds) is important for meteoroids with initial masses larger than a few grams, varying greatly from a few percent to almost 80 percent. We interpret this to signify that the most compact Geminids have masses between $\sim\!\!\!20\,\rm{g}$ and $200\,\rm{g}$. These Geminids lose mass primarily through regular ablation and reach the highest average pressure and the pressure for a $50\%$ mass loss (orange squares and green diamonds in Fig.~\ref{strength_stat}).

The strength proxy distribution shown in Fig.~\ref{strength_dist} corresponds to the trends seen in Fig.~\ref{maslost}, which shows the mass loss regimes. For small Geminids, most of the mass lost in initial fragmentation is caused by the erosion of grains released during fragmentation. The importance of erosion drops sharply for the moderate mass range of Geminids and ablation takes over. Therefore, the mass is released more gradually with respect to the aerodynamic pressure.

The mass evolution of the Geminids as a function of aerodynamic pressure is shown in Fig.~\ref{mass_evo}. Both ends of the mass spectrum (blue and orange to red) show more fluent mass changes with few to no gross fragmentations. The Geminids with moderate initial masses (cyan to green) exhibit more abrupt changes in total fragment mass (more frequent gross fragmentations with an immediate dust release). This is also consistent with Fig.~\ref{maslost} because the erosion of dust grains is a continuous process, and it is a prevailing mass loss process for small and large Geminids. Geminids with moderate initial masses mostly lose their mass by ablation and immediate dust release, as shown in Fig.~\ref{mass_evo} as a step function.

When we look at derived grain sizes in Fig.~\ref{grains}, we can see the modeled Geminids are made up of a mixture of various grain sizes ranging from about $20\,\rm{\mu m}$ to macroscopic parts or fragments of several millimeters in diameter. There does not seem to be a preferred fundamental grain size, as suggested in \citet{borovicka2010}. The minimum grain size was observed in Geminid~22, which exhibited a gross fragmentation at an altitude of $93.5\,\rm{km}$ above the ground, suddenly releasing about 1/3 of its original mass in very fine grains ranging from $7\,\rm{\mu m}$ to $90\,\rm{\mu m}$ in diameter. There is a possible observation bias: observations of faint meteors are usually taken with too small time resolution (30 and 61 fps in our case). Short flares caused by gross fragmentation can therefore easily be missed in faint meteors, as opposed to fireballs, for which we have high-time-resolution radiometric data.

The Geminids contain compact, non-fragmenting parts ranging in size from $1\rm{-}20\,\rm{mm}$ (blue disks in Fig.~\ref{grains}). These prevail in meteoroids weighing more than $20\,\rm{g}$ ($2.5\,\rm{cm}$ in diameter), but their total amount is variable. There are fewer of these in meteoroids up to $200\,\rm{g}$ and most of them are released in gross fragmentations. Above $200\,\rm{g}$ ($5\,\rm{cm}$), there is a large amount of these compact parts and they are released by erosion. These compact parts reach similar maximum dynamic pressures to those of smaller meteoroids (Fig.~\ref{strength_stat}). We primarily observe fine-grained material in meteoroids with a mass below $20\,\rm{g}$ and it is mostly eroded, while sometimes it can be released as immediate dust. Larger meteoroids contain a small amount of the fine-grained material.

A comparison of the semiautomatic and manual modeling shows that some model parameters are derived more robustly than others. Both methods yielded very similar values for the initial velocity, mass of the meteoroid, and erosion coefficients, as well as the upper grain mass limit. To the contrary, the ablation coefficient and the lower grain mass limit sometimes differ by a factor of two. This can be explained by the fact that ablation was scarcely observed before the onset of erosion in the modeled faint Geminids and the lower grain mass limit is not well constrained by the video data.

A power-law trend is seen in the mass-velocity plot in Fig.~\ref{mass_velocity}. The initial velocity decreases as the meteoroid mass increases. This suggests that Geminids of different sizes have different orbits in the Solar System. Of course, our sample size is very small and a more substantial dataset could change this perspective.

There are several caveats to bear in mind. For Geminids smaller than $\sim\!\!2\,\rm{g}$, the minimum grain mass in our model (for immediate or eroded dust) is somewhat arbitrary. We tested this for Geminid 16. We were able to fit the data with similar quality for minimum grain masses ranging from $10^{-11}\,\rm{kg}$ to $10^{-8}\,\rm{kg}$. There were only negligible differences in the light curve fit and virtually no differences in the dynamics fit. Throughout the modeling, our default value for the minimum grain mass was set to $10^{-11}\,\rm{kg}$, and we relied on the modeling process to choose the appropriate grain mass. This is true for high-fidelity radiometric light curve data in fireballs, but it is doubtful for smaller meteoroids. For these, we have light curves derived from the video data with 61 fps.

All the derived masses rely on the luminous efficiency functions we applied. We are quite confident in the results for the large Geminids, for which we used the \citet{revelle2001} function modified by \citet{borovicka2020b}. At the other end of the mass spectrum, we used the function of \citet{pecina1983}; however, we note that it only depends on meteor velocity and not its mass. Moreover, it is extrapolated to the moderate velocities of Geminids from much lower velocities, for which it was derived. For these meteors, the \citet{revelle2001} function gives too high a luminous efficiency, so it is difficult to reconcile our model with both the light curve and dynamical data using this luminous efficiency. In the moderate-mass range ($1\rm{-}40\,\rm{g}$), we mostly used the \citet{pecina1983} function, but there is probably a transition from lower to higher luminous efficiency values somewhere between $20\rm{-}40\,\rm{g}$. Ultimately, the horizontal axis of many of our plots is distorted to some degree due to the mass uncertainty.

\section{Conclusions}\label{conclusions}
We modeled the fragmentation cascades of 39 Geminid fireballs and fainter meteors using a semiautomatic procedure based on genetic algorithms. Some of them (1--9) were taken from a previous study \citep{henych2024} with minor adjustments, and some faint Geminids (32--39) were also modeled manually to validate our method. The second part of this paper comprises a numerical modeling of the atmospheric thermal stresses acting on model meteoroids with the masses, velocities, and trajectory slopes of the observed Geminids, as well as material properties similar to those of carbonaceous chondrites. We list our main conclusions below.

Based on the correlation between the derived minimum fragmentation pressure of the observed Geminids and the mass dependence of failure due to thermal stress, we conclude that thermal stress initiates the fragmentation of the Geminids before they start to ablate and can be observed. The mechanical forces that crush the Geminids only become significant later in the lower layers of the atmosphere.

The mass-loss regime of Geminids depends on their initial mass. The gradual release of dust grains (erosion) is important for Geminids with $m<20\,\rm{g}$ and for those with $m>200\,\rm{g}$. We note that small Geminids release relatively small dust grains, while large Geminids can release macroscopic particles. These two processes are formally described as erosion in the model of \citet{borovicka2020b}. Gross fragmentation with immediate dust release plays a significant role for  Geminids with moderate initial masses ($40\rm{-}200\,\rm{g}$). The most compact Geminids are those with $20\,{\rm g}<m<200\,\rm{g}$.

We observed no preferential grain size in modeled Geminids; the minimum grain size is usually $20\,\rm{\mu m}$. However, this result is somewhat model-dependent for faint Geminids. Larger Geminids ($m>200\,\rm{g}$) contain a large amount of compact, non-fragmenting parts ranging in size from $1$ to $20\,\rm{mm}$.
Geminids of different sizes are likely to have different orbits in the Solar System. Their initial velocity decreases as their initial mass increases.

We derived the bulk densities of ten fainter Geminids using two different approaches. Their bulk densities vary widely, but they are generally higher than $1800\,\rm{kg\,m^{-3}}$ (with one exception). For one larger meteoroid, the bulk density is equal to the assumed grain density ($3000\,\rm{kg\,m^{-3}}$).

\section*{Data availability}
The global physical classification (PE, Pf and KB) in a single file, light curve data (*.lc), dynamics data (*.inp), and radiometric data (*.rlc) for all 39 meteors used in the present study are available on Zenodo (\href{https://doi.org/10.5281/zenodo.17868507}{https://doi.org/10.5281/zenodo.17868507}).

The light curve data, radiometric, and dynamics data contain the following quantities. Relative time is time measured from an arbitrary initial time instant common to all stations. The absolute brightness of a meteor is measured in magnitudes (calculated from the measured brightness and calibrated for a distance of $100\,\rm{km}$ from the station), and the magnitude error represents the uncertainty of the measured brightness. Length is the distance along a meteor's trajectory measured from its beginning position, measured typically from the closest station. Height is the altitude above ground level, calculated using triangulation of the meteor's trajectory observed from several stations. The length and height in INP files are measured values for a given time mark or video frame. These data were used for the modeling of the dynamics. The length and height in the LC file are smoothed values from a preliminary time-length fit that does not consider fragmentation. These data are only provided for plotting brightness as a function of length or height.

\begin{acknowledgements}
We thank Galina O. Ryabova for her comments and questions that helped us improve the presentation of our results. This research was supported by grant no. 24-10143S from the Czech Science Foundation. Computations were performed on the OASA and VIRGO clusters at the Astronomical Institute of the Czech Academy of Sciences. This research has made use of NASA's Astrophysics Data System.
\end{acknowledgements}

\bibliographystyle{aa}
\bibliography{gem_fragile}
\end{document}